# Multiple Andreev reflections in diffusive SNS structures

Rafael Taboryski, Jonatan Kutchinsky, and Jørn Bindslev Hansen

Department of Physics, Technical University of Denmark, building 309, DK-2800
Lyngby, Denmark

Morten Wildt, Claus B. Sørensen, and Poul Erik Lindelof

Niels Bohr Institute, University of Copenhagen, Universitetsparken 5, DK-2100
Copenhagen Ø, Denmark

**Abstract**

We report new measurements on sup-gap energy structure originating from multiple Andreev reflections in mesoscopic SNS junctions. The junctions were fabricated in a planar geometry with high transparency superconducting contacts of Al deposited on highly diffusive and surface δ-doped $n^{++}$-GaAs. For samples with a normal GaAs region of active length $0.3\mu m$ the Josephson effect with a maximal supercurrent $I_c=3\mu A$ at $T=237mK$ was observed. The sub-gap structure was observed as a series of local minima in the differential resistance at dc bias voltages $V=\pm 2\Delta/(ne)$ with $n=1,2,4$ i.e. only the even sub-gap positions. While at $V=\pm 2\Delta/e$ ($n=1$) only one dip is observed, the $n=2$, and the $n=4$ sub-gap structures each consists of two separate dips in the differential resistance. The mutual spacing of these two dips is independent of temperature, and the mutual spacing of the $n=4$ dips is half of the spacing of the $n=2$ dips. The voltage bias positions of the sub-gap differential resistance minima coincide with the maxima in the oscillation amplitude when a magnetic field is applied in an interferometer configuration, where one of the superconducting electrodes has been replaced by a flux sensitive open loop.

**1. Introduction**

The Andreev reflection is a second order quantum mechanical process by which an electron-like particle incident on the superconductor with a quasiparticle excitation energy $\varepsilon$ above the Fermi energy, may be transmitted as a Cooper pair in the superconductor, if a hole-like particle with energy $-\varepsilon$ is retroreflected along the path of the incoming electron [1]. For a superconductor-oxide-normal metal (S-I-N) interface, with a very thin oxide layer, or for an superconductor-semiconductor interface with a negligible Schottky barrier this leads to an increased conductance, which is seen as the so-called excess current at high voltage bias ($V > 2\Delta/e$). The sub-gap structure (SGS) is normally observed in Superconductor-Normal-conductor-Superconductor (SNS) junctions with peaks in the conductance at bias voltages $V = 2\Delta/ne$ with $n = 1,2,3....$ This relation simply expresses the condition for maximum charge transfer for multiple Andreev reflections given the number of traversals of the normal region. The SGS was given a semiclassical explanation in Ref.[2]. However, the semiclassical model completely neglects scattering and the influence of coherent states in the normal region. Recently also coherent transport in short SNS junctions has been taken into account [3]. Meanwhile, considerable progress has been made in the understanding of the diffusive transport in normal-conductors in contact with superconducting electrodes that carry no current, but induce a strong non-equilibrium state in the normal-conductor via the superconducting proximity effect [4]. Here, highly involved theory based on the Usadel equations and Greens functions techniques is employed.

The superconducting energy gap Δ defines the voltage scale on which the sub-gap structure is observed. This also sets the relevant ranges of Josephson frequencies $2eV/\hbar$. The present models for SGS in SNS junctions all rely on the assumption, that the transit time across the normal region is short compared to the inverse Josephson frequency at all voltages. In diffusive normal-conductors with diffusion constant $D$, the average transit time over the length $L$ is $L^2/D$. The junction must then be short compared to the length $L_\Delta = (\hbar D/\Delta)^{1/2}$. The coherent charge transport across the SN interface rely on the Andreev reflection mechanism, which couples the single electron states at energies $\mu_s \pm \varepsilon$, where $\mu_s$ is the chemical potential in the superconductor. In a diffusive system, the energy difference $2\varepsilon$ between the states is the limiting factor for how far from the SN interface the states can remain phase-coherent. A pair of Andreev reflected electron and hole like quasiparticles moving from the interface will build up a phase-difference $\delta\phi = 2\varepsilon t/\hbar$ over the time $t$. During this time the particles will diffuse an average distance $L = (Dt)^{1/2}$. This defines the relevant energy scale for dephasing of the state pair $\varepsilon_c = \hbar D/L^2$, also called the Thouless energy. If phase-breaking mechanisms are present during the transport, the phase-breaking diffusion length $\ell_\phi$ will enter the correlation energy $\varepsilon_c = \hbar D(1/L^2 - 1/\ell^2)$. This is sketched in Fig.1. The states can only remain coherent in the energy window $\mu_s \pm \varepsilon_c$, which decreases away from the interface. Gueron et al. [5] have shown that this window contains an induced quasi-gap in the density of states. The supercurrent coherence length in the normal conductor $\xi_N = \left(2\pi k_B T/(\hbar D) + 1/\ell_\phi^2\right)^{-1/2}$ is defined by the condition that the states out to this length should be coherent in the energy window $\pm 2\pi k_B T$. The critical current $I_c$ is thus determined by states with energies of the order of the thermal energy. For a long SNS junction with a normal region longer than $L_\Delta$

Andreev reflections will couple the same pair of states on both sides, only if the two superconductors are at the same potential. This gives rise to bound Andreev states in the coherent window. However, if a voltage is applied, the multiple Andreev states are no longer bound and a strong non-equilibrium quasiparticle distribution is present in the normal region. In addition, the coherent states are still restricted to coherent windows of widths $2\varepsilon_c$, but since the potentials on both sides are no longer aligned, there will be no overlap between the coherent windows emerging from each superconducting electrode. This is the main difference between long and short junctions. In a short junction, the entire range of involved energies can carry coherent states.

A gap in the understanding is the so-called fully out of equilibrium situation, where a finite voltage is applied between superconducting electrodes to a long diffusive normal-conductor. However, despite the apparently overwhelming theoretical complexity of this situation we will demonstrate, that the experimental results are surprisingly clear. In this paper we present new data on the SGS in long and highly diffusive SNS junctions.

**2. Experimental techniques**

In the experiment the normal-conductor consisted of a 200 *nm* thick degenerate GaAs layer grown in a VARIAN molecular beam epitaxy chamber on an undoped GaAs substrate. The 200 *nm* of GaAs were doped with Si to $4.4 \times 10^{18} cm^{-3}$ and capped with five Si monolayers (δ-doping) seperated by 2.5 *nm* of GaAs. Each of the δ-doped layers contained $5 \times 10^{13}$ Si atoms per $cm^3$. The purpose of these layers were to decrease the thickness of the Schottky barrier at the S-Sm interface, which was formed by depositing 200 *nm* of Al in-situ after the substrate had been allowed to cool down. This procedure produced a highly transparent superconductor-semiconductor interface with a measured contact resistivity between Al and GaAs for the best material (HCØ296) of $\rho_c = 8 \times 10^{-13} \Omega m^2$. An estimate of the barrier transmission coefficient based on the excess current in devices made from this material gives $T_n = 0.5$. On the other hand, an estimate based on the contact resistivity gives $T_n = (3\rho_c/(\ell\rho) + 1/2)^{-1} = 0.15$, where $\ell \approx 50 nm$ and $\rho \approx 8 \times 10^{-6} \Omega m$ is the mean free path and resistivity of the semiconductor. This discrepancy may be explained if the effective contact area is roughly a factor of 0.4 smaller than the nominal contact area. $T_n = 0.15$ is then a lower bound for the transmission coefficient. The contact resistivity depends crucially on the MBE growth parameters. The shown differential resistance traces in Fig.2 are measured on samples cut from different wafers all with exactly the same layer configuration, but with differences in the detailed growth parameters. This results in huge variations of the contact resistivities. However, samples cut from the same wafer all had very similar characteristics. All data in the rest of the paper refer to samples cut from the best wafer, HCØ296.

The GaAs had a carrier density of $n = 4.8 \times 10^{18} cm^{-3}$ and a diffusion constant $D = 0.016 m^2/s$. For an electrode separation of $L = 1\mu m$ this gives a Thouless energy of $\varepsilon_c = \hbar D/L^2 = 10.4 \mu eV$. The phase-breaking diffusion length was measured in an independent experiment by mean of the weak localisation effect [6] and gave $\ell_\phi \approx 5\mu m$ at the base temperature of the cryostat. The resulting base temperature normal metal coherence length is then $\xi_N = (2\pi k_B T/(\hbar D) + 1/\ell_\phi^2)^{-1/2} \approx 0.28 \mu m$. A consequence of the planar geometry is that at each Al/GaAs interface the current will flow from the superconducting Al film to the resistive GaAs over a typical decay length for the current density $\ell_n = (d\rho_c/\rho_{GaAs})^{1/2} \approx 0.1\mu m$, where $d$ and $\rho_{GaAs}$ are the thickness and the resistivity of the GaAs respectively. The Al film had a superconducting transition temperature close to the bulk value $T_c$=1.2 K. For individual devices the superconducting energy gap ranged from Δ=170 μeV to 178μV clustered around the bulk value of 175 μeV. The processing of the MBE grown wafer started with the etching of an 18 *μm* wide mesa structure in the Al and GaAs layers with separate current and voltage contacts for four point measurements. Then the SNS devices were made by removing Al in selected areas by use of electron beam lithography with standard PMMA resist and wet etching [7]. The devices were either shaped as simple Josephson junction type of slits with lengths *L*, or flux-sensitive interferometers where one of the Al electrodes was shaped as an open loop, as seen in the inset of Fig.5.

The electrical measurements of the samples were carried out in a $^3$He cryostat with a base temperature of $0.235 mK$. The measurement leads were filtered with $0.5m$ cold THERMOCOAX filters providing high frequency power attenuation of $12.5 \times (f[Ghz])^{1/2} dB$ [8] and room temperature π filters giving approximately 20 *dB* attenuation at 700 *kHz*. The leads between the cold and R.T. filters were pair twisted to avoid power line pick up. The *I/V* and *dV/dI* vs. *V* characteristics of the devices were measured by applying a dc bias current superimposed a low frequency ac modulation current. The ac voltage response was measured with a lock in amplifier referenced to the modulation frequency. The overall resistance of the devices was of the order of $1-10\Omega$ depending on geometry. The ac current amplitude was kept as low as possible to gain optimal resolution in the dc measurement without compromising the signal to noise ratio of the ac measurement. A sensible compromise was found by using approximately $200 nA$ rms ac excitation amplitude.

In the interferometer experiment a magnetic field was applied from a superconducting magnet driven with a high resolution current source. In experiments where the magnetic field was not swept, it was carefully zeroed to within $3.3\mu T$ corresponding to about one tenth of a flux-quantum penetrating the effective flux sensitive area obtained when flux expulsion from the superconducting areas due to the Meissner effect have been taken into account.

**3. Results and discussion**

For the sample with the shortest SNS junction $L \approx 0.3\mu m$, both the DC and AC Josephson effect with a maximal supercurrent of $I_c = 2.9\mu A$ at $T = 237 mK$ was observed. The DC effect is shown in Fig.3. The supercurrent could be clearly resolved up to $T = 0.6K$. However, a detailed temperature dependence of $I_c$ cannot be extracted from the data. In Fig.4 the traces for a $L = 1.1\mu m$ junction are shown. The sharp dip at zero bias is an indication of a not yet fully developed or noise rounded supercurrent. The presence of a fully developed supercurrent for the $0.3\mu m$ sample but not for the $1.1\mu m$ sample is consistent with the calculated coherence length of $0.28\mu m$. The magnetic field dependence of the *I-V* characteristics is of the well known "Fraunhofer" type (not shown), indicating a homogenous current distribution across the junction.

The differential resistance measurements shown in Figures 4 and 5 represent the key data of this paper. The sub-gap structure is seen as a series of local minima in the differential resistance. In the existing theories the SGS is predicted at DC bias voltages $V = \pm 2\Delta/(ne)$, where *n* takes on all positive integer values. As shown in Fig.6 for the short sample the SGS is observed at $n = 1,2,4$ i.e. only at even sub-gap positions. The $n=4$ feature is only clearly resolved in the $0.3\mu m$ sample, although a detailed measurement and a subtraction of the background do scarcely reveal the $n=4$ structure in the $1.1\mu m$ sample. This can be explained by the sum of the four traversals of the normal region becoming comparable to $\ell_\phi$ for the $n=4$ structure in the longer sample. The experimental fact that only even SGS is seen may indicate that phase-coherence between the electron-like quasiparticles and the Andreev reflected holes plays an important role for the SGS. The even SGS positions all reflect situations where Andreev reflected quasiparticles pair wise emerge at the same energy, and thus provide phase-coupling of the two superconducting electrodes via coherent energy windows. The $n=2$ SGS thus corresponds to a situation where the coherent energy window emerging from one electrode probes the quasiparticle density of states singularity at $\varepsilon=\Delta$ in the other electrode. Quasiparticles injected into the normal-conductor at $V=\pm\Delta/e$ will remain phase-coherent with their Andreev reflected counterparts. The short and the long junction qualitatively posses the same features in the differential resistance traces, despite that the Thouless energy and consequently the coherent energy window is much wider in short junction, namely $\varepsilon_c \approx 112\mu eV$ for the $0.3\mu m$ junction compared to $\varepsilon_c \approx 8\mu eV$ for the $1.1\mu m$ junction.

In the interferometer experiment shown in Fig.7 the current carried by coherent states is distinguished from the background current by application of a magnetic field perpendicular to the loop plane [9]. The magnetic flux threading the loop imposes a phase-difference across the slit between the arms of the loop. The observed oscillations disappear below experimental resolution with increasing bias, as the voltage exceeds $\varepsilon_c/e \approx 10.4\mu V$. The oscillations reappear at higher bias around the superconducting energy-gap $V=\Delta/e$ corresponding to the SGS for $n=2$. The $n=4$

SGS could not be resolved in the interferometer experiment where the distance between the split electrode and the counter-electrode was roughly $1\mu m$. At $V = \pm 2\Delta/e$ merely no oscillations are observed, despite the huge dip in the differential resistance at this bias position. This reflects that at this bias no Andreev reflection is necessary to obtain quasiparticle transfer between the two electrodes, and thus no electron-hole coherence is established. The tiny peak in the oscillation amplitude at this bias position has a different origin and can be atributed to a very weak oscillating magnetic field dependence of the energy gap. The resulting oscillations are thus phase-shifted by $180^0$ as the voltage passes the $V = 2\Delta/e$ minima. The oscillations at all other bias positions are in phase. This experiment clearly indicates that phase-coherent transport plays an important role at the SGS bias positions.

Both in the differential resistance traces and in the oscillation amplitude a pronounced splitting of the SGS was observed. While at $V = \pm 2\Delta/e$ ($n = 1$) only one dip is observed, the $n = 2$, and the $n = 4$ SGS each consists of two pronounced separate dips in the differential resistance. Moreover in Fig.5 an additional very sharp dip between the B1 and B2 dips is also seen. This extra dip was only observed in the short junction. For the short junction measurements in Fig.5 the mutual spacing of the two sets of local minima (B1,B2) and (C1,C2) is independent of temperature. The spacing of the $n = 2$ dips is roughly $58\mu V$, while the spacing of the $n = 4$ dips is $28\mu V$ i.e. about half the value of spacing of the $n = 2$ dips. For the long junction the splitting of the SGS amounts to roughly $30\mu V$ and again about half this value for the weak $n = 4$ SGS splitting. The origin of this splitting is not understood and calls for further theoretical investigation.

## 3. Conclusion

In conclusion, we have observed the Josephson effect and sub-gap energy structure originating from multiple Andreev reflections in highly diffusive Al/GaAs/Al SNS junctions. The SGS could be related to the phase-coherent transport observed in a flux sensitive interferometer at finite voltage bias $V = 2\Delta/(2e)$. A pronounced splitting of the SGS was observed, but is at present not understood.

**Acknowledgements**

This work was supported by the Danish Technical Research Council, the III-V Nanolab, and the CNAST center program.

**Figure Captions**

**Fig.1**

Decay of coherence as a function of distance from the SN interface. Energy is shown in the y-direction. The normal metal coherence length $\xi_N$ and the phase-breaking diffusion length $\ell_\phi$ are also shown.

**Fig.2**

The DC Josephson effect for the $0.3\mu m$ sample. The inset shows the temperature of the maximal supercurrent.

**Fig.3**

Differential resistance vs. applied DC bias voltage for samples cut from three different wafers, with nominally identical layer configurations, but grown under slightly different growth conditions.

**Fig.4**

The differential resistance vs. applied DC bias voltage for the $1.1\mu m$ junction for different temperatures. All traces except base temperature trace have been offset for clarity.

**Fig.5**

The differential resistance vs. applied DC bias voltage for the $0.3\mu m$ junction for different temperatures. The local minima in the differential resistance associated with energy gap positions are marked. All traces except base temperature trace have been offset for clarity.

**Fig.6**

The positions of the local minima marked in Fig.4 plotted vs. temperature. Also shown are the BCS temperature dependence of the energy gap and relevant sub-gap energies. The experimental values of $2\Delta/e$ are slightly suppressed due to Joule heating of the carriers. The used $T_c = 1.175K$ which is slightly lower than the bulk value of for Al $1.196K$ is fitted to the data, and the value of the energy gap $\Delta(0)$ is chosen correspondingly lower than the bulk value.

**Fig.7**

Upper left: Oscillations in the differential resistance taken at the highest V=Δ/e peak shown in the middle panel. Upper right: sample layout. The oscillation period correspond to one h/2e flux quantum per cycle. Middle and bottom: Comparison between the differential resistance and the relative oscillation amplitude, both plotted as a function of bias voltage.

**Taboryski et al. Fig.1**

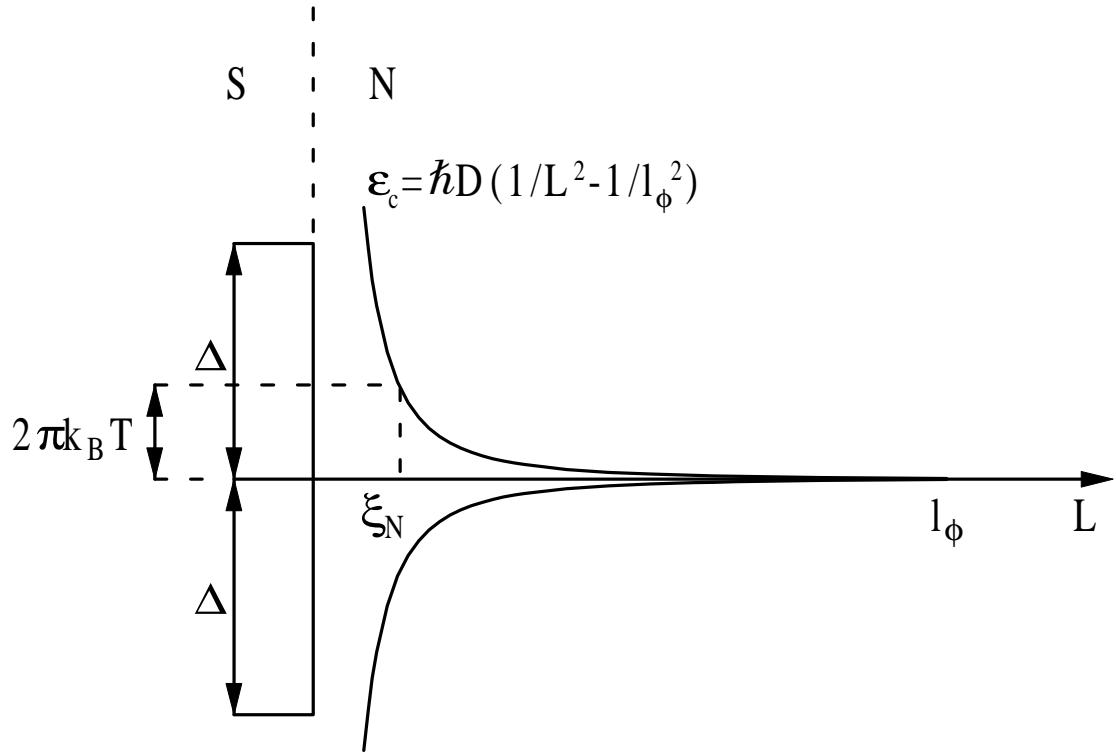

Taboryski et al. Fig.2

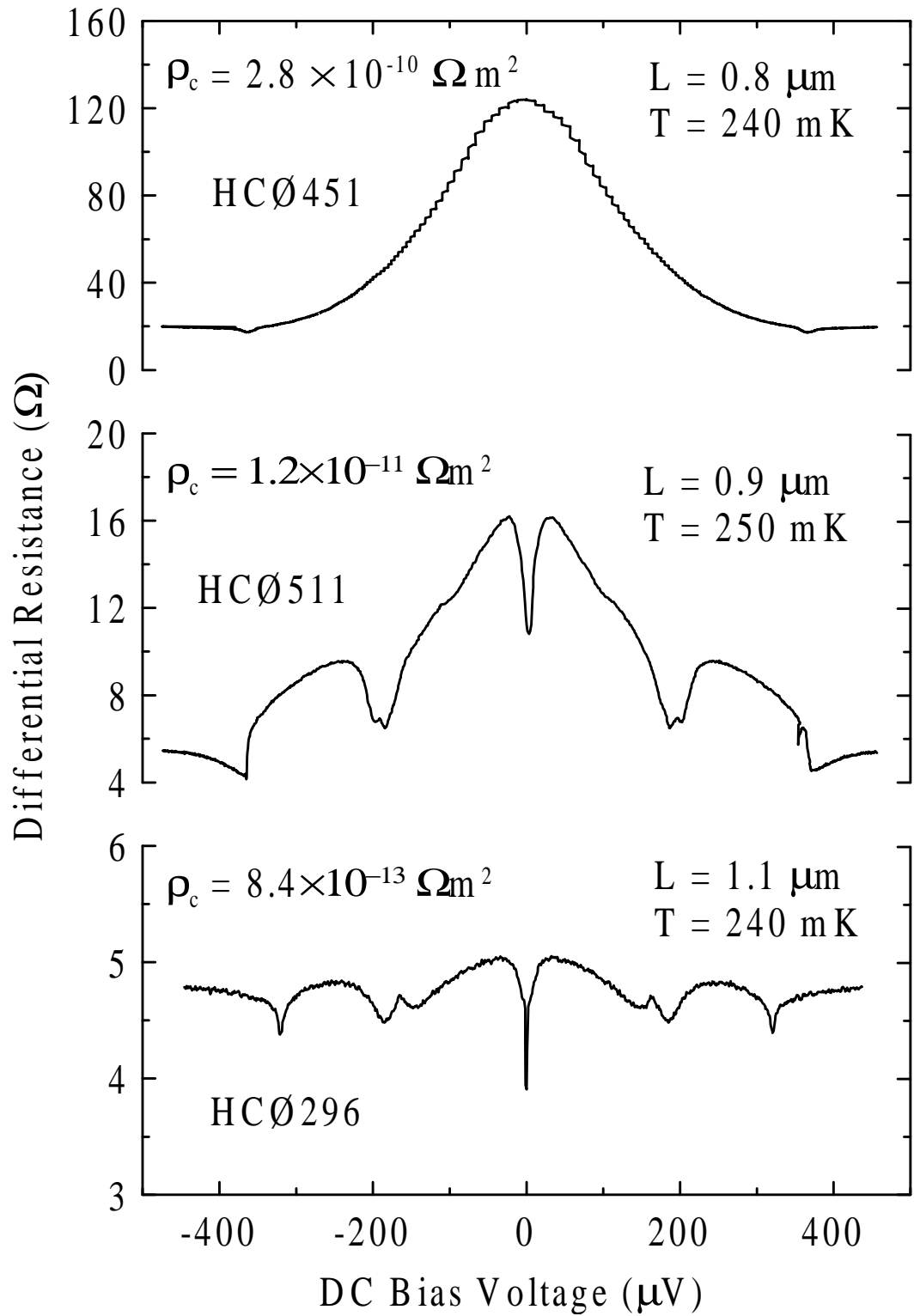

**Taboryski et al. Fig.3**

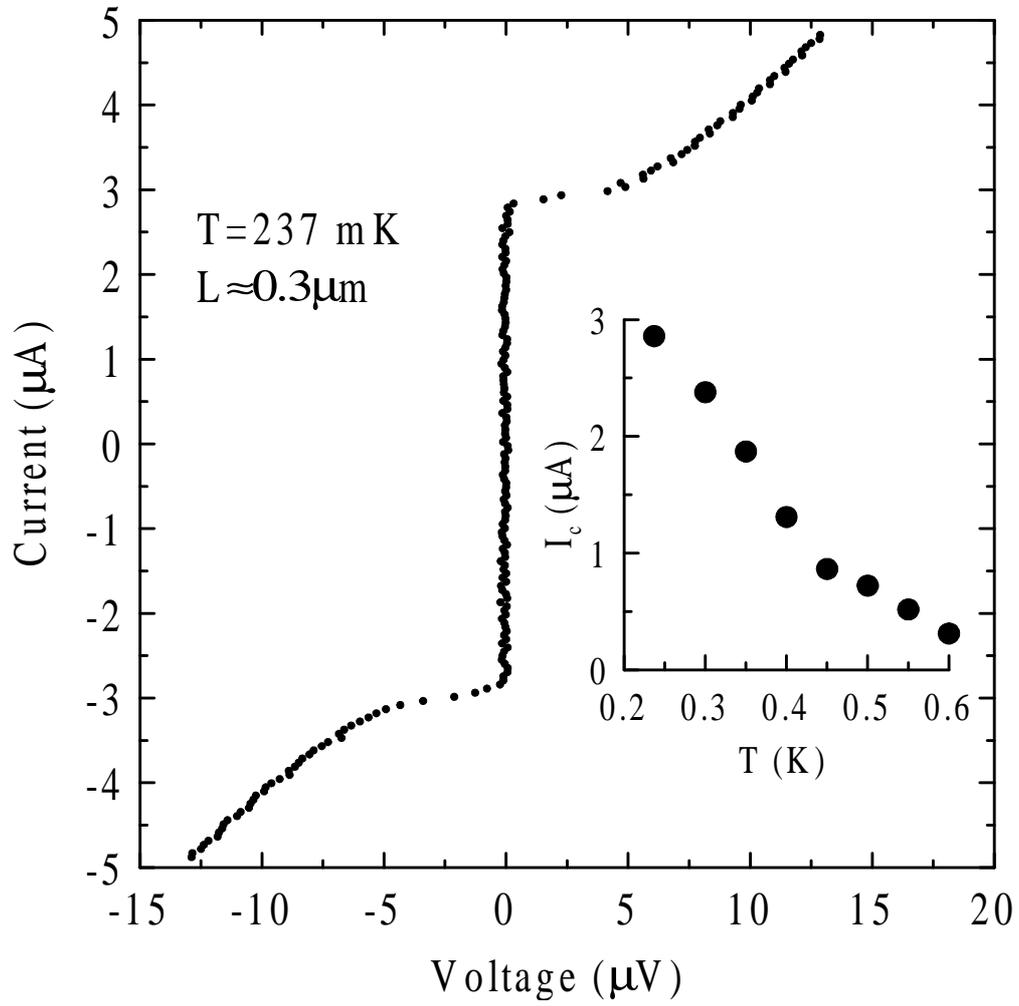

**Taboryski et al. Fig.4**

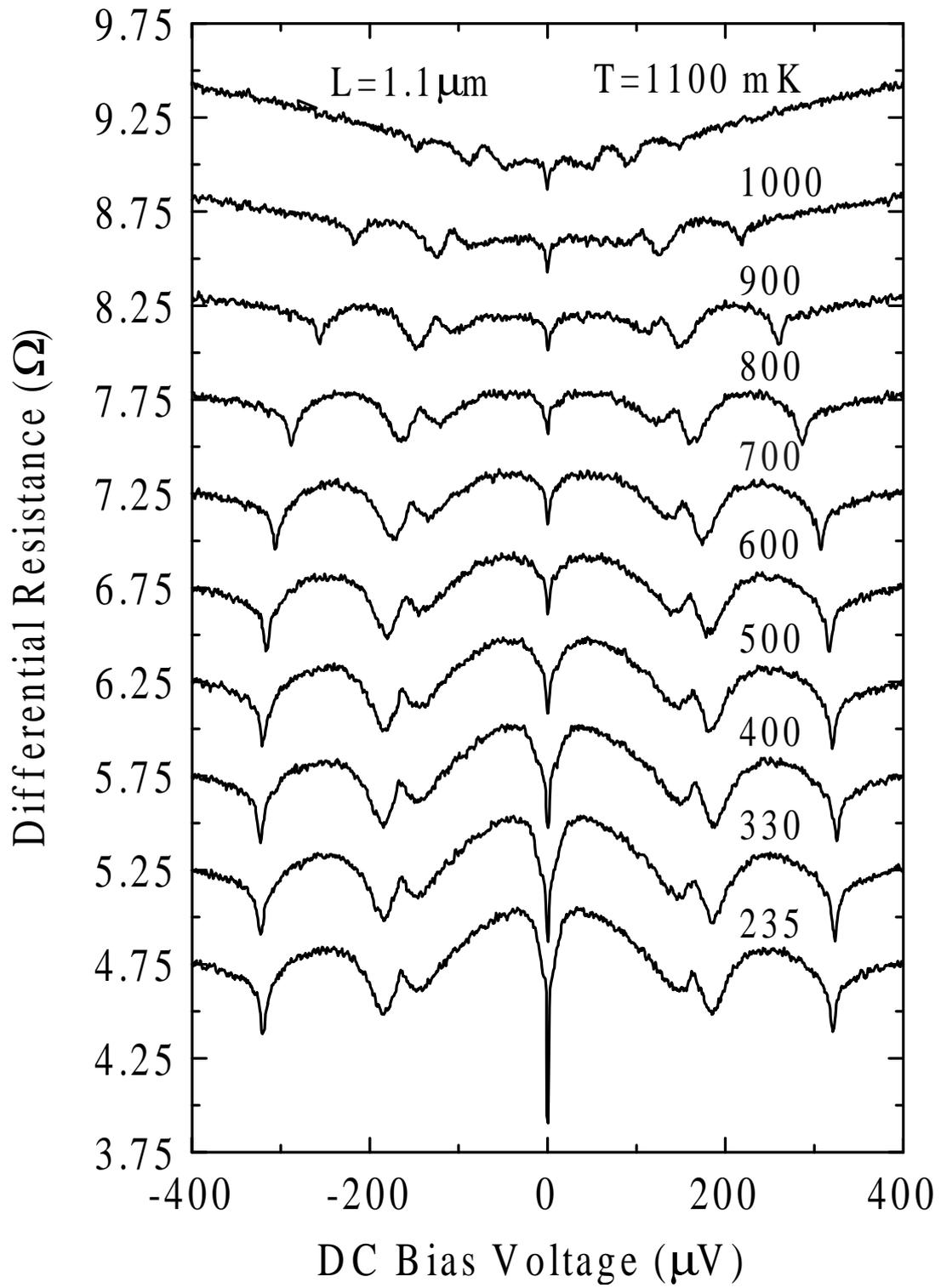

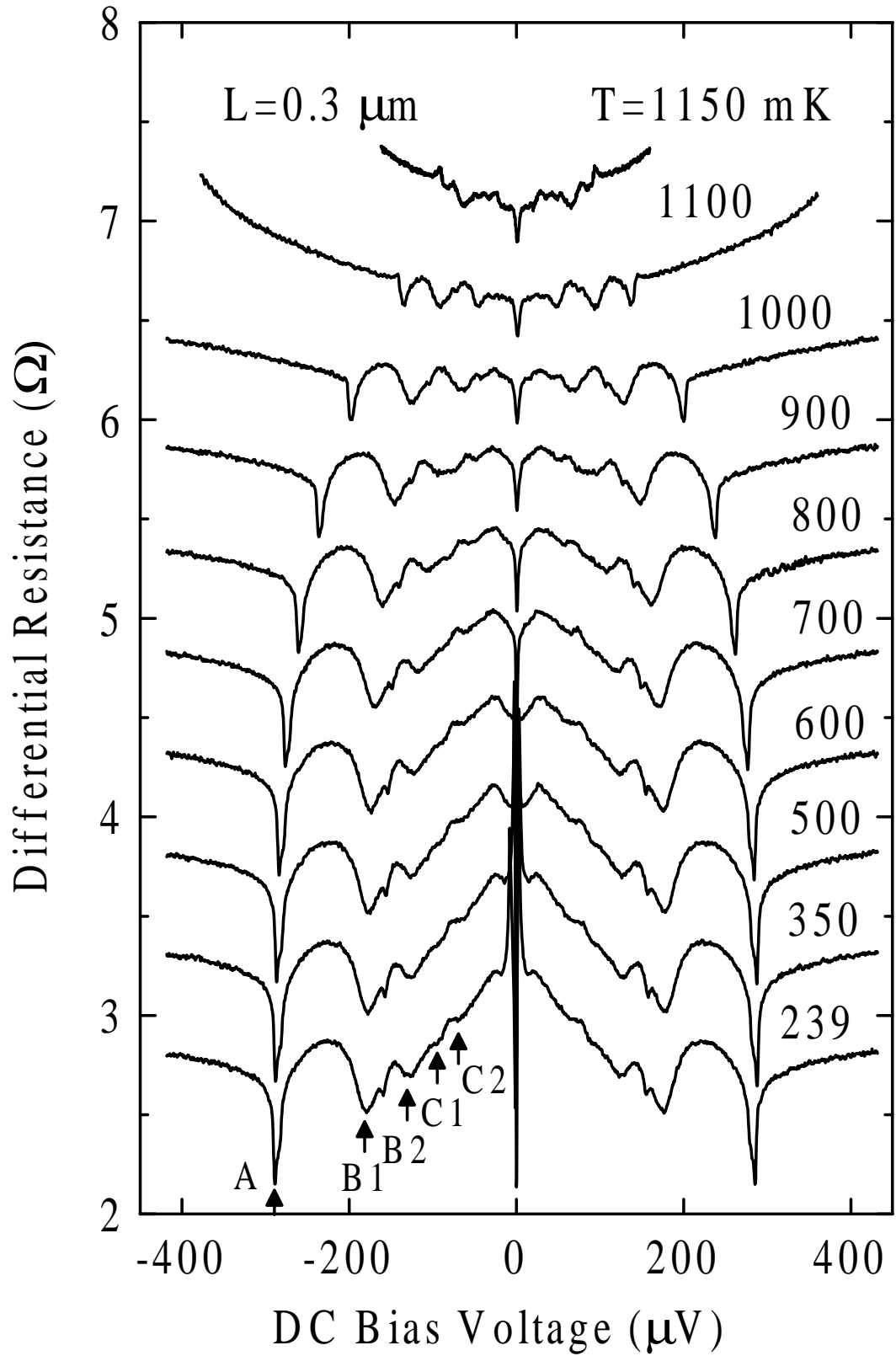

**Taboryski et al. Fig.6**

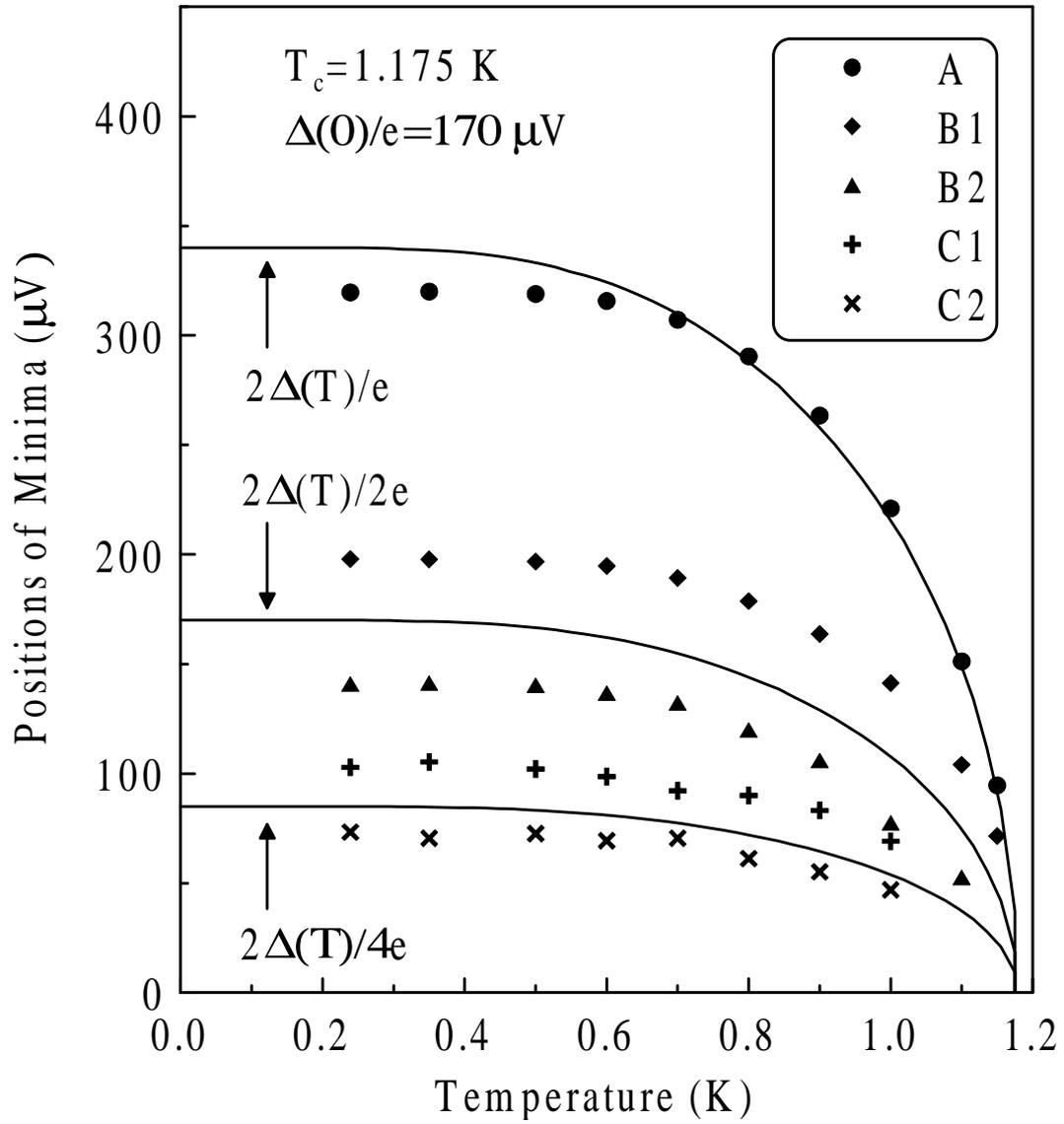

**Taboryski et al Fig.7**

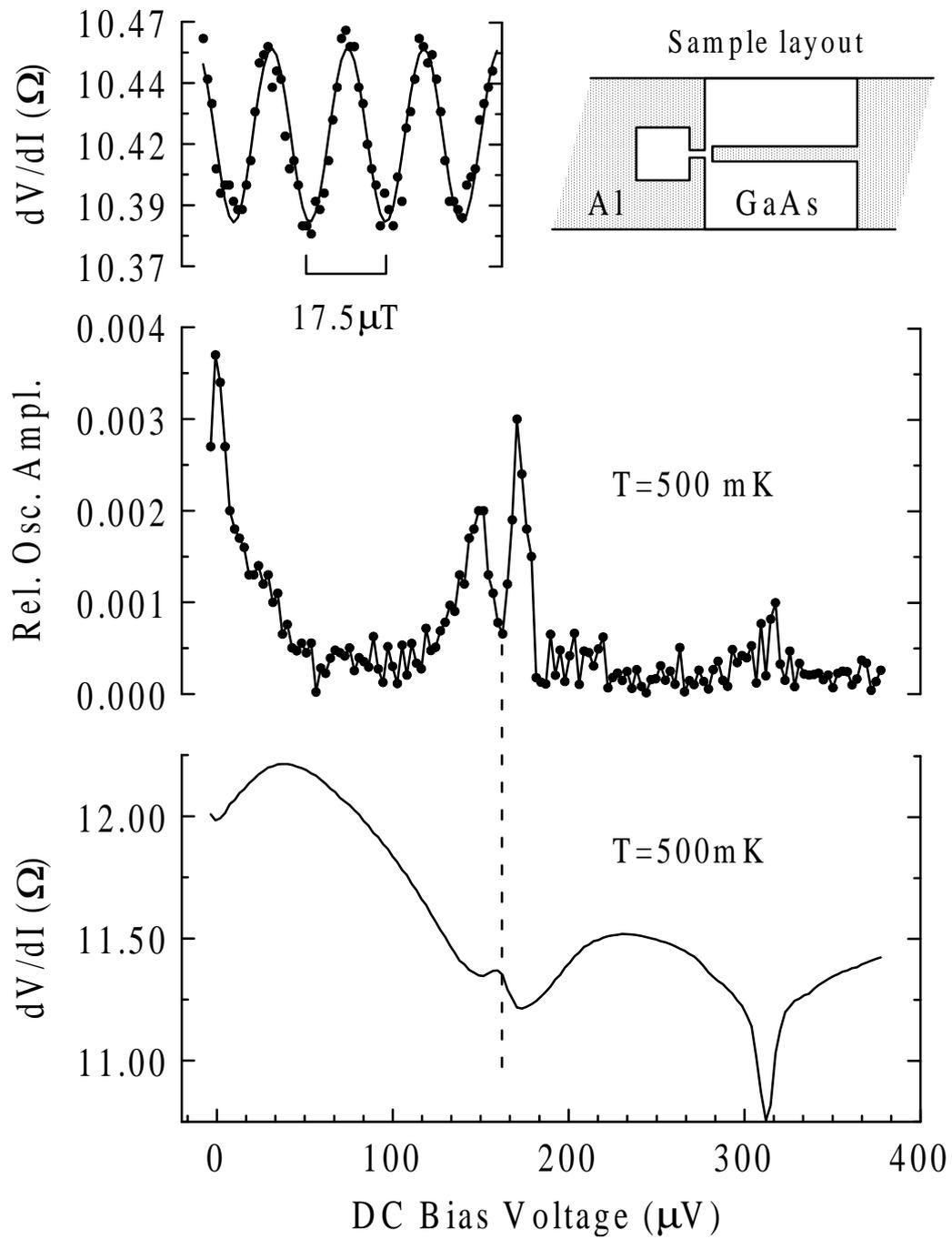